\documentclass{article}
\usepackage[utf8]{inputenc}
\usepackage[english]{babel}

\usepackage{graphicx}
\usepackage{epsfig}            
\usepackage{pstricks}          

\usepackage{fancyhdr}
\usepackage[verbose,a4paper,tmargin=30mm,bmargin=37.5mm,lmargin=30mm,rmargin=30mm]{geometry}

\usepackage{helvet}

\usepackage{mathptmx} 
\usepackage[T1]{fontenc}

\usepackage{rotating}

\usepackage{amssymb}       
\usepackage{amsmath}       

\usepackage{natbib} 

\usepackage{multirow}

\usepackage{verbatim}          
\usepackage{subcaption}
\definecolor{METblue}{cmyk}{0.85,0,0.2,0.2}

\usepackage{titlesec}

\usepackage{url}
\usepackage{authblk}

\title{High-resolution monthly precipitation climatologies over Norway: assessment of spatial interpolation methods}
\date{}

\author[1]{Alice Crespi}
\author[2]{Cristian Lussana}
\author[3]{Michele Brunetti}
\author[2]{Andreas Dobler}
\author[1,3]{Maurizio Maugeri}
\author[2]{Ole Einar Tveito}
\affil[1]{Department of Environmental Science and Policy, Università degli Studi di Milano, Italy}
\affil[2]{The Norwegian Meteorological Institute, Oslo, Norway}
\affil[3]{Istituto di Scienze dell’Atmosfera e del Clima, CNR, Bologna, Italy}
\date{}         

\begin{document}
\maketitle
\bibliographystyle{agufull04}

\section*{\hspace{17mm}Abstract}
Monthly precipitation climatologies at 1 km resolution have been produced over the Norwegian mainland for 1981-2010.
The observed station normals are interpolated over a regular grid by applying a multi-linear local regression kriging (MLRK).
The statistical method aims at modeling the influence of the main geographical features, such as: latitude, longitude, elevation and sea nearness on the precipitation field at a local scale.
The MLRK is composed of two steps, (i) a background precipitation field is computed through a multi-linear local regression based on the geographical information, then (ii) a kriging interpolation is applied to adjust the field so to better fit the station residuals (i.e., the difference between the observed normals and the background field).
The interpolation accuracy is evaluated by reconstructing the station normals with a leave-one-out approach and by comparing the model performances with those of other interpolation methods.\\
The poor observation coverage over remote and mountainous regions in Norway has motivated us to consider precipitation fields produced by numerical models. In fact, numerical model output provides a reference field for the evaluation of MLRK that is not dependent on the station density, though it is not as accurate as the observations.
Specifically, a regional climate simulation with a resolution of 2.5 km, based on the dynamical downscaling of the global reanalysis ERAInterim and available for the time period 2003-2016, has been used.\\
In addition to the MLRK validation, the combination of numerical model fields and observed climatologies has been investigated and the results show that this integrated approach provides more accurate high-resolution climatologies over the Country. The numerical background introduces valuable information over remote and mountainous regions, whereas in-situ data correct for model biases.

\section{Introduction}
High-resolution gridded datasets of surface precipitation represent important supporting tools for decision makers in a wide range of fields such as energy production, management and conservation of natural sources as well as in operative meteorology. 
In particular, due to their primary role in water supply, hydrological balance and hydro-power production, gridded datasets of monthly precipitation are essential products for Norway and they are required in both climatological and hydrological applications.\\
In this document, a new version of 1981-2010 monthly precipitation climatologies at 1 km resolution over Norway is presented. We aim at representing small scale features in the climatologies, where the definition of small scale depends on the local spatial distribution of observations.
The precipitation fields are based on in-situ observations only and statistical methods have been applied for the spatial interpolation.
The observations we use have been quality checked and the raw data were considered, without applying any correction for the wind-induced undercatch.
The statistical method we implemented takes into account the strong relationship between rainfall and surface geographical features, that is well documented by the scientific literature (e.g. \cite{DALY94}; \cite{MASSFREI14}). More precisely, monthly precipitation climatologies of Norway are interpolated by the \textit{multi-linear local regression kriging} (MLRK) in which gridded normals are obtained by a two-step approach.
First, a precipitation field is obtained by means of a multi-linear local regression. For each grid point, the neighboring stations are selected and a multi-linear function of their geographical features (elevation, latitude, longitude and sea nearness) is fitted to the correspondent precipitation normals. 
Second, the precipitation field obtained in the first step is used as the background field for a residual kriging procedure (\cite{GOOV2000}), where the residuals between the observed normals and the background values at station locations are spatially interpolated over the high-resolution grid.\\
The observation network plays a crucial role in the determination of the quality of the climatologies (\cite{hofstra2008comparison}; \cite{HOF09}).
In dense station areas we get more accurate estimates while in data sparse areas the final results are affected by a higher uncertainty.
The Norwegian station network is strongly uneven with higher station density in the South, a sparser network in the North and a poor coverage over the mountainous areas, especially at higher elevations.
As a consequence, interpolation schemes, such as those proposed in \cite{DALY02} and \cite{cres18} modeling precipitation on a very local scale and using elevation as the main predictor are more likely to produce unrealistic patterns over those complex and scarcely observed areas.
On the contrary, MLRK takes into account a larger number of geographical predictors and it takes into account larger spatial scales so that it is more robust with respect to inhomogeneities in the spatial distribution of observations.\\
The monthly climatologies have been evaluated through cross-validation methods. In particular, we have considered two domains, one in the north and one in the south in order to study the influence of station density and different climatic conditions on the model performances.\\
Our study focuses also on the comparison between observational gridded datasets and numerical model output.
Gridded numerical model data for hourly precipitation is provided by the climate model version of HARMONIE (version cy38h1.2), a seamless NWP model framework developed and used by several national meteorological services. HARMONIE includes a set of different physics packages adapted for different horizontal resolutions. For the high-resolution, convection permitting simulations in this case, the model has been set-up with AROME physics \citep{Seity2001Arome} and the SURFEX surface scheme \citep{Masson2013SURFEX}. The climate runs (hereafter referred to as "HCLIM-AROME") have been carried out within the HARMONIE script system, covering the period July 2003 to December 2016 on a 2.5 km grid over the Norwegian main land. More details on the climate model can be found in \cite{Lind2016HCLIM},  references therein and on \emph{https://www.hirlam.org/trac/wiki/HarmonieClimate}.
Because the numerical model does not include measurements from the network of rain-gauges, the output is an independent information that can be used to evaluate observational datasets. 
Furthermore, the effective resolution of the precipitation field obtained by numerical models can be considered rather uniform over the spatial domain.
For this reason, the numerical model output can be used as a reference to assess the relative increase of uncertainty in the observational gridded dataset while moving from data dense areas towards data sparse areas.\\
The promising results obtained during our study, motivated us to investigate the combination of HCLIM-AROME data and in-situ observations. In principle, such a combination integrates the accuracy of the observations and the fine-scale realistic patterns of the convection resolving numerical model.
The MLRK scheme has been modified so to include HCLIM-AROME monthly averaged precipitation fields as the background for the residual kriging.
This combination turned out to improve the quality of average monthly precipitation fields over the time period 2003-2016 that is covered by HCLIM-AROME, if compared to the observational dataset.
Therefore, the proposed combination method can be used to obtain more accurate monthly precipitation normals over Norway.

\section{Observation data}
\begin{figure}
\centering
\includegraphics{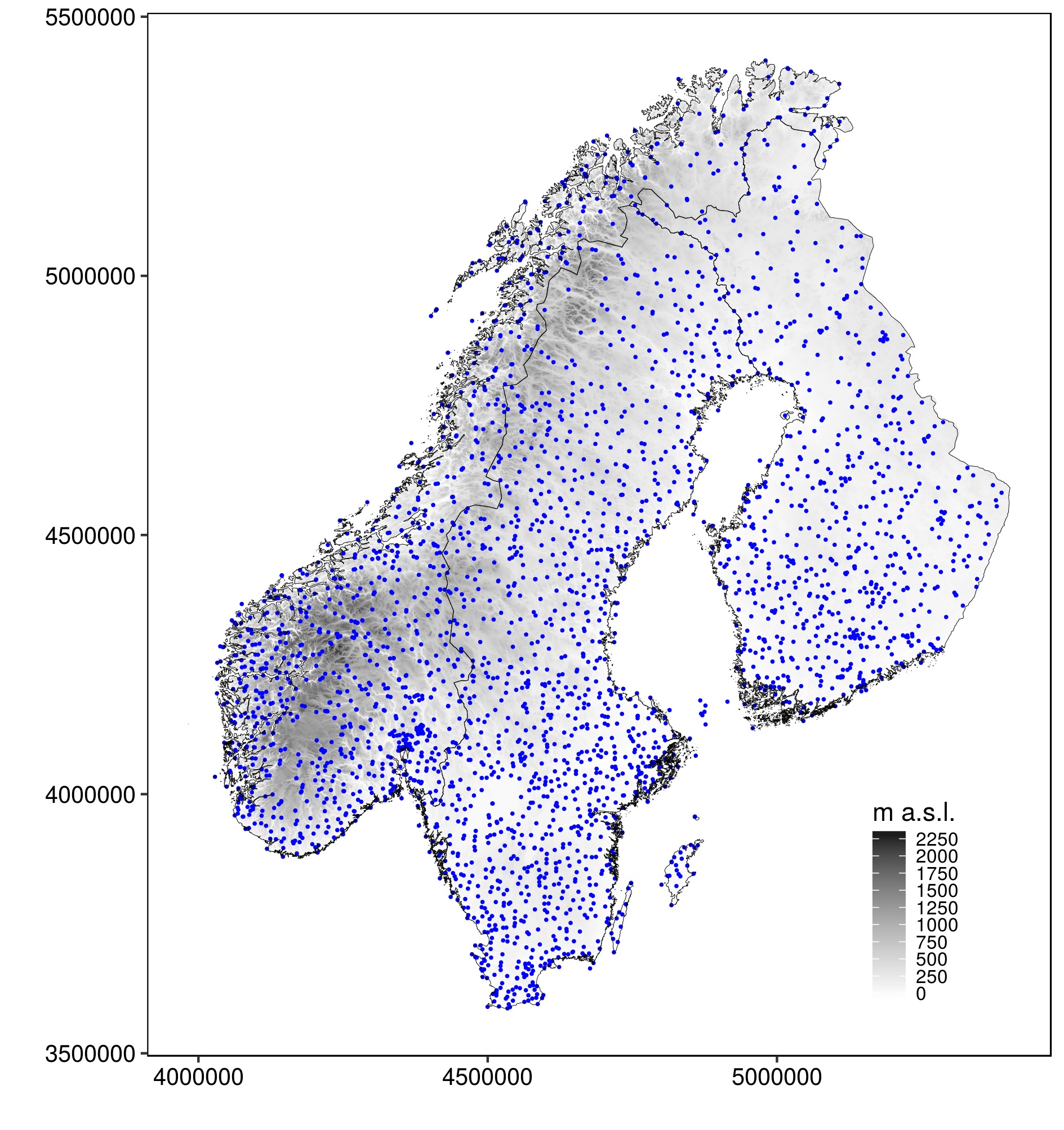}
\caption{Spatial distribution of available stations over Scandinavian peninsula.}
\label{fig:stat_cov}
\end{figure}

\begin{figure}[h]
\includegraphics{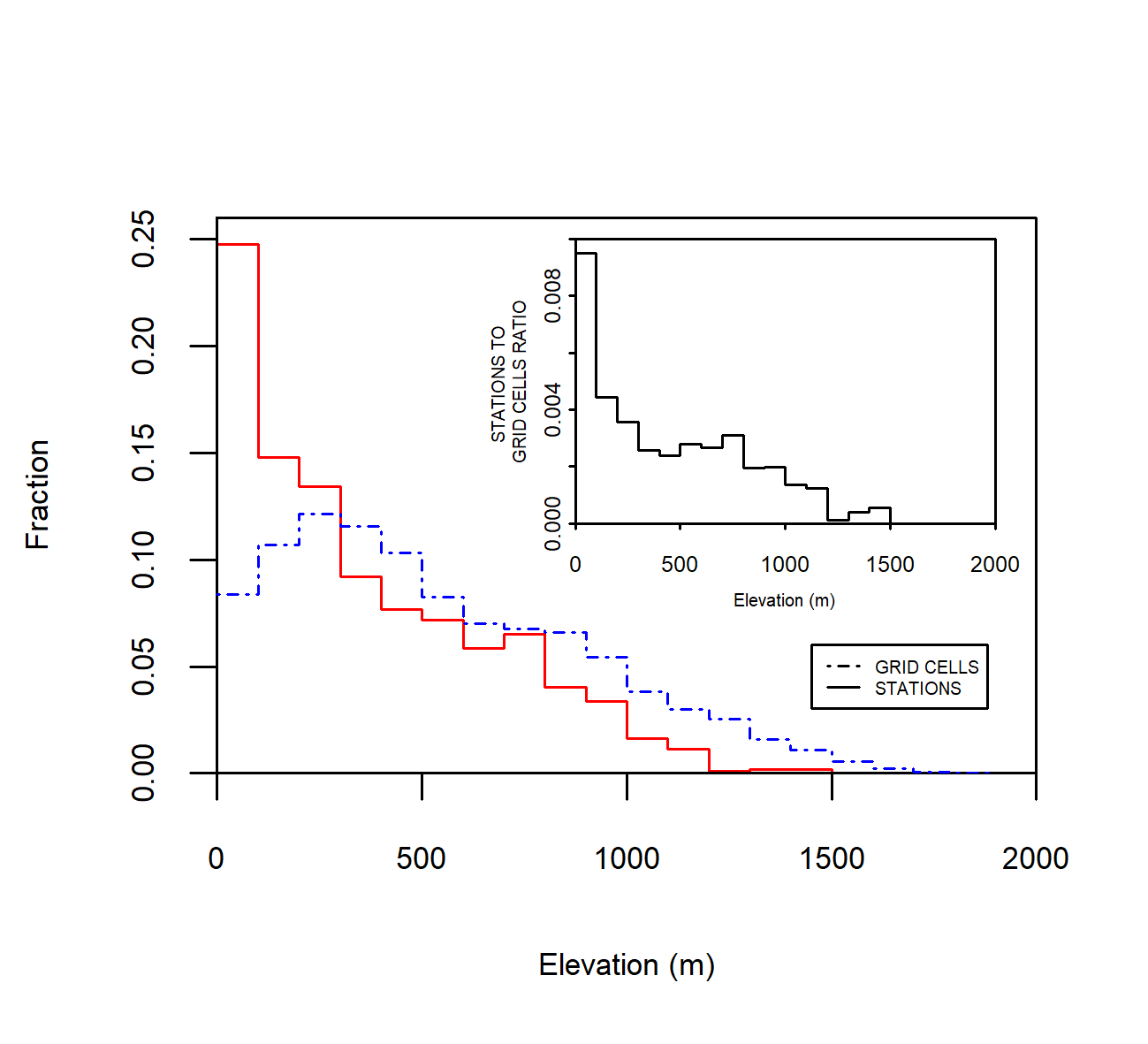}
\centering
\caption{Vertical distribution of the 1043 Norwegian stations (solid line) compared to the grid cell elevation distribution (dashed line) over Norway. The inset box shows the vertical distribution of the stations to grid cells ratio.}
\label{fig:vert_dist}
\end{figure}

The database considered for the construction of 1981-2010 
monthly precipitation climatologies for Norway includes 
more than 5000 daily precipitation series covering 
Fennoscandia (Norway, Sweden and Finland) and neighboring 
countries. Daily precipitation series were retrieved from 
ECA\&D archive \url{https://www.ecad.eu//dailydata/index.php} and the MET Norway Climate database (KDVH). More precisely, ECA\&D series relative to 
Norwegian sites were replaced by MET stations if they were 
closer than 100 m (radial distance) and 1 m (altitude 
difference), while those MET series without a 
correspondence in ECA\&D archive were added to the database as 
new stations. When necessary, ECA\&D daily records were 
shifted in order to represent the accumulated 
precipitation between 06 UTC of previous day to 06 UTC of 
current day, which is the definition for daily 
precipitation in KDVH series. Moreover, metadata 
containing geographical information (latitude, longitude 
and elevation) are available for each series.\\ The database was 
subjected to a careful quality control procedure aiming at 
detecting duplicate records, spurious data entries and 
correcting erroneous station locations. In order to 
identify duplicates each series was compared to the ones 
available in a radius of 15 km and common data in every 
year (only values greater than 1 mm were considered) were 
checked for similarities. If the fraction of identical 
entries was higher than 70\% for a certain year, the two 
series were 
put into correspondence. Single periods of duplicate 
values at different site locations are likely to be due to 
erroneous data assignments and they were invalidated, while 
in case of greater overlapping periods only the longest 
and most continuous series was retained. In addition, in 
order to avoid different data referring to the same 
geographical location, stations showing equal coordinates 
were checked and their metadata corrected by means of 
geographical information provided by the respective national 
meteorological services or Google Earth.\\
After these procedures, monthly totals for each station 
were computed and their data quality was evaluated on a monthly basis by comparing measured records with a 
simulated series obtained by means on neighboring 
stations. More precisely each monthly entry of each 
station (\textit{test}) was simulated by considering ten reference series. They were selected among the closest stations which have enough 
common data (at least 9) for the month under reconstruction with the test 
series and in which the entry to simulate is available. The monthly value of test series 
(\(p_{test_m})\) was computed from the corresponding datum 
of each of the ten reference series (\(p_{ref_m,i})\) as 
follows:
\begin{equation}
\label{Eq.1}
\tilde{p}_{test_m,i}=p_{ref_m,i} \cdot\frac{\bar{p}_{test_m}}{\bar{p}_{ref_m,i}} \quad\quad  (i = 1,..., 10)
\end{equation}
where \(\bar{p}_{test_m}\) and \(\bar{p}_{test_m,i}\) are the test’s and reference series’ averages for the considered month over their period of common data. The final estimation of \(p_{test_m}\) was finally obtained considering the median of the ten values. High reconstruction errors, in terms of mean absolute error (MAE) and root mean square error (RMSE), allowed to point out and remove single problematic periods in a record or to discard stations whose measured values completely mismatched the simulated ones.\\
After removing gross data errors, the daily series underwent a gap-filling procedure aiming at maximizing the number and length of monthly data series available for climatological purposes. The procedure was managed only if daily gaps were contained in a month with valid entries for more than 2/3 of days and for at least 50\% of days in a 5-month period centered on it. For each gap the stations in a radius of 50 km from the reconstructed one were considered and correlation computed. A daily gap was reconstructed by considering the value of the series showing the highest correlation (minimum threshold of 0.7) and valid entries in common for at least 50\% of the 5-month period. The daily entry of reference series was reported to the mean daily value of the station to fill by using as rescaling factor the ratio between their daily averages over the common data period.\\
After fulfillment, monthly precipitation series were computed again for each station. Whenever daily data were still missing, the corresponding monthly total was not computed. On average, the filling procedure allowed to reconstruct about 5\% of missing daily data for each station, which corresponded to a benefit in terms of monthly total precipitation computation of about 15\%.
Monthly series relative to very close sites with contiguous and homogeneous data were merged in order to maximize the data coverage over the period chosen as reference for climatologies, while stations with less than 10 years of available monthly data, also after gap-filling and merging activities, were definitely discarded from the database. \\After these activities, 1981-2010 monthly precipitation normals were computed for each series and, whenever this period was partially or completely unavailable, missing months were reconstructed following the same method used for quality-check (Eq. \ref{Eq.1}) but taking as final value the weighted mean of the ten reference station estimates rather than their median. Since a remarkable fraction of stations (52\%) had more than 30\% of missing data in the reference period, this procedure allowed to prevent monthly normals being biased by a lower fraction of monthly data entering in the average computation.\\
The resulting database available for the construction of Norwegian 1981-2010 high-resolution precipitation climatologies is composed by 3226 observation sites covering Norway and surrounding countries as shown in figure \ref{fig:stat_cov}. 1043 stations are located in Norway, even though they are unevenly distributed over the domain: data coverage is generally higher in the South, especially around the Oslo fjord, while observation availability decreases significantly in the North especially over inland. Data availability gets even lower if high-elevation areas are considered as pointed out in figures \ref{fig:stat_cov} and \ref{fig:vert_dist}, where distribution of available stations for elevation range is depicted. The significant fraction of grid cells at high-elevation not covered by stations, whose distribution strongly decrease especially above 1000 m a.s.l., suggests that data coverage is not completely representative of study domain features.  


\section{Methods\label{sec:methods}}

\subsection{Definition of interpolation grid\label{sec:smooth}}
Precipitation climatologies for Norway are gridded on a smoothed version of 1 km x 1 km resolution Digital Elevation Model (DEM) expressed in Lambert Azimuthal Equal-Area projection. This DEM version is retained by rescaling the 30 arc-second resolution GTOPO30 (\textit{USGS,1996}) on a 1 km x 1 km grid and assigning to each cell an elevation obtained as a weighted average of elevations of surrounding cells. The station weight is expressed as a Gaussian function decreasing to 0.5 at a distance of 3 km from the cell itself. The smoothing degree is defined by varying the distance at which weights are halved and choosing that minimizing the model errors (see below for details on model errors). Smoothing of orographic details is performed in order to adapt terrain representation to the actual scale at which precipitation-orography interactions are supposed to occur \citep{FORESTI2011}. Main geographical parameters are computed for each cell (\(x,y)\) of smoothed DEM version:

\begin{itemize}
\item slope orientation (\textit{facet})
\begin{equation}
\label{Eq.2}
facet(x,y)= \begin{cases}
    \arccos\left(\frac{\frac{\partial{z}}{\partial{x}}}{|\vec{\nabla} z|}\right) & \text{if } \left(\frac{\partial{z}}{\partial{x}}\right)>0\\
  2\pi-\arccos\left(\frac{\frac{\partial{z}}{\partial{x}}}{|\vec{\nabla} z|}\right) & \text{if } \left(\frac{\partial{z}}{\partial{x}}\right)<0
\end{cases}
\end{equation}
\item slope steepness
\begin{equation}
\label{Eq.3}
Sl(x,y) = |\vec{\nabla} z|=\sqrt{\left(\frac{\partial{z}}{\partial{x}}\right)^2+\left(\frac{\partial{z}}{\partial{y}}\right)^2}
\end{equation}
\item distance from the sea is based on \textit{cross-distance} computation which combines horizontal distance and vertical gradients taking into account the presence of orographic obstacles.
\end{itemize}
Geographical information for each station site, including smoothed elevation, are finally extracted from the nearest DEM cell and used in interpolation procedures described in next sections.
\subsection{Interpolation model: multi-linear local regression kriging\label{sec:interp}}
High-resolution 1981-2010 monthly precipitation climatologies for Norway are computed by performing \textit{Multi-linear Local Regression Kriging (MLRK)} at each cell of smoothed DEM. More precisely, this procedure is composed by two main steps:
\begin{enumerate}
\item \textit{multi-linear regression} of precipitation \textit{versus} chosen geographical predictors is performed at each station site \textit{i} and residual between observed and estimated values is computed:
\begin{equation}
\label{Eq.4}
\epsilon^i=p^i-\tilde{p}^i=p^i - \sum_{j=0}^{n} \alpha_j^i \cdot q_j^i
\end{equation}
 where \textit{n} is the number of predictors, \(\alpha\)  is the vector of regression coefficients and \(q^i\) is the vector of predictors at target site. 
Latitude, longitude, elevation and distance from the sea are chosen as regression predictors while \(\alpha\) are estimated for each month and for each station by least-square method considering precipitation normals and geographical features of all sample sites within 100 km from the station under consideration. If less than 100 stations are available for regression within this distance, searching radius is incremented of 10 km until the minimum threshold is reached. Both the searching radius and the number of stations entering in the regression were defined by the minimization of model errors (see below for details on model errors).
\item \textit{ordinary kriging} (OK) of station residuals. In this case, experimental semi-variogram is constructed by setting bin width to 15 km and maximum pair distance to 300 km and by using exponential model as fitting curve.
\end{enumerate}
Finally, precipitation at each grid cell is estimated:
\begin{equation}
\label{Eq.4}
p(x,y)=\sum_{j=0}^{n} \alpha_j(x,y) \cdot q_j(x,y) + \mathbf{k^T}(x,y) \cdot \epsilon
\end{equation}
where \(\mathbf{k}(x,y)\) is the vector of kriging weights for cell \((x,y)\) and \(\alpha(x,y)\) are the regression coefficients estimated at grid point \((x,y)\) by considering its geographical features \(q_j(x,y)\) and a minimum threshold of 100 stations within 100 km from it.

\subsection{The combination of HCLIM-AROME and observations \label{sub:HCLIM}}

\begin{figure}
\centering
\includegraphics[width=15cm,height=16cm]{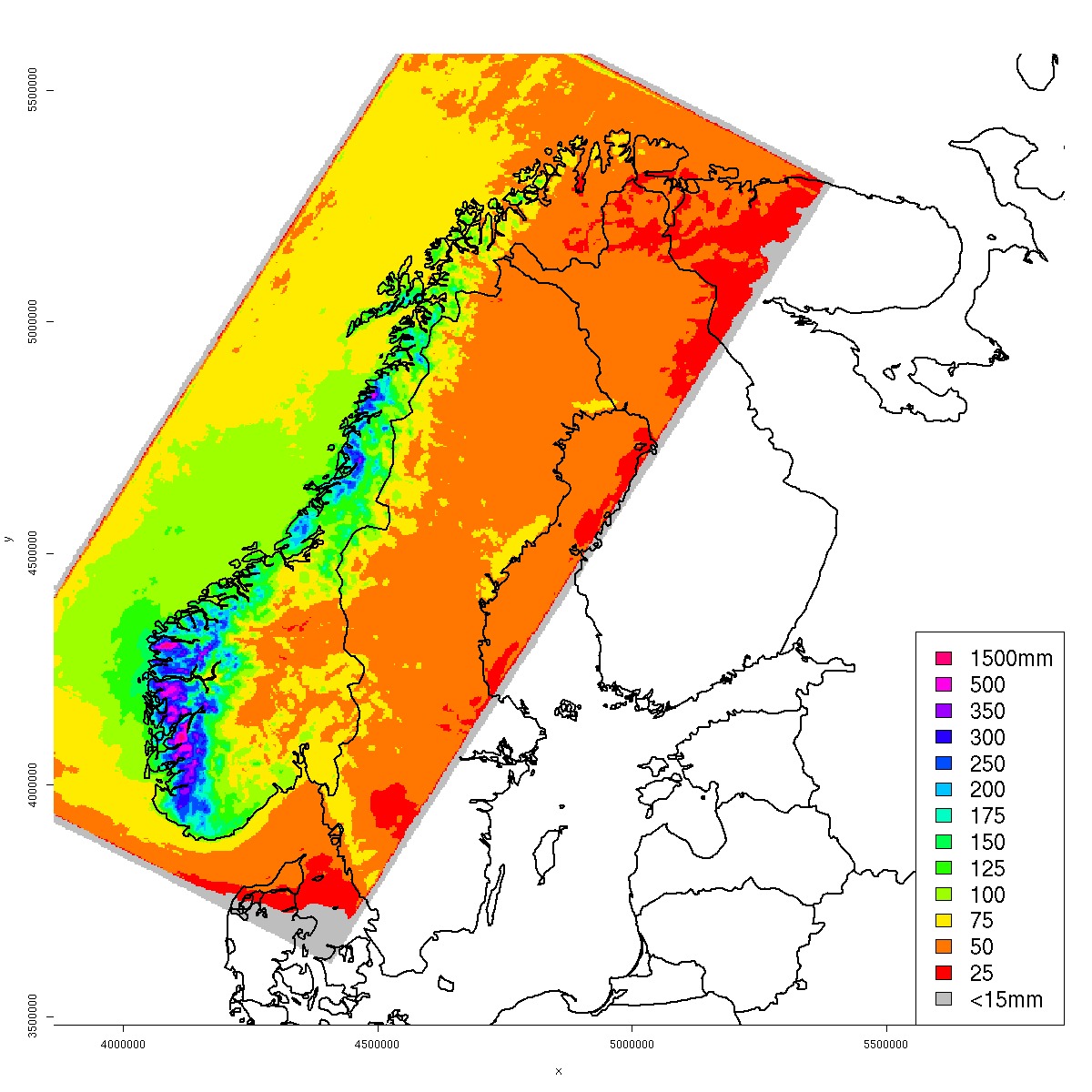}
\caption{HCLIM-AROME 2.5 km resolution field of mean Januar precipitation 2004-2016.}
\label{fig:HCLIMclima}
\end{figure}

In order to evaluate the robustness of MLRK and its limits in handling with areas with sparse station networks, statistical model outcomes were compared with precipitation fields provided by a numerical model and a new interpolation approach combining numerical model results with station observations was tested. To this aim, monthly total precipitation series produced by HCLIM-AROME on a 2.5 x 2.5 km grid covering Norway and a portion of Sweden (figure \ref{fig:HCLIMclima}) for 2003-2016 were considered. 
Monthly normals for each cell of 2.5 km grid were computed and the 12 resulting fields were then downscaled to the target 1 km resolution grid in ETRS89-LAEA projection by means of bilinear interpolation routine provided by the \textit{raster} package in R software.  In order to cover the same reference period, monthly normals for 2003-2016 interval were computed for each station site and used to provide MLRK gridded fields to be compared with HCLIM-AROME ones.  Moreover, high-resolution monthly precipitation climatologies were computed  by means of a kriging-based procedure combining HCLIM-AROME numerical model predictions with station observations (HCLIM+KR). In HCLIM+KR, HCLIM-AROME monthly field is substituted to the outcomes of multiple linear regression in the first step of MLRK and it is used as background field in kriging interpolation. More precisely, the residuals between station normals and the values extracted from the closest HCLIM-AROME grid cell are modeled by OK and results used to correct initial numerical field. Final precipitation estimation at each grid cell \((x,y)\) is obtained as the sum of HCLIM-AROME value at the point and corresponding kriging correcting coefficients: 
\begin{equation}
\label{Eq.6}
p(x,y)= p_{hclim}(x,y) + \mathbf{k_{hclim}^T}(x,y) \cdot \epsilon_{hclim}
\end{equation}

As in MLRK scheme, experimental variogram is constructed by setting bin width and maximum pair distance to 15 km and 300 km, respectively, and theoretical variogram is obtained by means of exponential fit.

\section{Results}
\subsection{Model validation}
MLRK performances were evaluated by comparing 1981-2010 monthly precipitation normals of the 1043 Norwegian stations with the corresponding values estimated by model in leave-one-out approach (LOO), i.e. by removing the value of station under reconstruction in order to avoid self-influence. More precisely, due to computational time reasons, the covariance matrix was computed from the whole dataset, while the kriging weight of the station under consideration was set to 0 and the remaining station weights were re-normalized. 
\begin{table}[h]
\centering
 \begin{tabular}{||c c c c c||} 
 \hline
 MONTH & BIAS & MAE & RMSE  & \(R^2\) \\ [0.5ex] 
 \hline\hline
 1 & 0.8 & 16.0 & 25.1  & 0.9\\ 
 2 & 0.6 & 12.9 & 20.4 & 0.9\\
 3 & 0.6 & 12.2 & 19.0 & 0.9\\
 4 & 0.3 & 8.5 & 12.9 & 0.9 \\
 5 & 0.1 & 7.1 & 10.8 & 0.8\\ 
 6 & 0.0 & 7.2 & 10.7 & 0.8\\ 
 7 & 0.1 & 7.9 & 11.8 & 0.8\\ 
 8 & 0.1 & 9.5 & 14.6 & 0.9\\
 9 & 0.2 & 12.4 & 20.4 & 0.9 \\
 10 & 0.5 & 14.4 & 23.0 & 0.9\\ 
 11 & 0.7 & 13.9 & 21.8 & 0.9\\  
 12 & 0.7 & 15.2 & 24.3 & 0.9 \\ 
 \hline
\end{tabular}
\caption{Accuracy of MLRK monthly climatologies obtained from the leave-one-out validation for the 1043 stations in Norway. All values are expressed in mm and BIAS is defined as difference between simulation and observation.}
\label{tab:MLRK}
\end{table}

In table \ref{tab:MLRK} model errors obtained for all months are listed in terms of BIAS, MAE, RMSE and \(R^2\) as average over all validation stations. The highest errors are obtained for winter months, especially in January, when the greatest precipitation values occur over Norway, especially over the mountainous region close to south-western coast. It is noteworthy that, as depicted in figure \ref{fig:vert_dist}, observation coverage decreases significantly on high-elevation areas contributing to limit model ability in fully capture rainfall gradients over orographically complex regions. 
However, \(R^2\) is above 0.8 for all months, suggesting a generally good agreement between modeled and measured values. Low values of monthly BIAS, even if systematically positive, give evidence of no significant under or overestimation of station normals by model reconstruction. In addition, the influence of station density on MLRK accuracy was investigated by comparing LOO model errors obtained for Northern sites (above Trondheim, 63\(^{\circ}\)18' N) and Southern sites (below 63\(^{\circ}\)18' N) separately. The reconstruction errors, both in absolute and relative terms, turn out to be 10\% higher for the Northern stations than for the Southern ones for all months. This discrepancy could be partly explained by the remarkable difference in data coverage characterizing the two parts of the country. \\
In order to better verify the robustness of kriging applied to residuals of a multiple regression approach and the suitability of chosen regression predictors, we compared the results provided by other interpolation approaches. More precisely, we took into account 1) an OK on the residuals of a local regression of precipitation \textit{versus} elevation and 2) a \textit{local weighted linear precipitation-elevation regression} (LWLR, \cite{cres18}) in which stations selected for the linear fit are weighted according to their nearness and orographic similarities (elevation, sea distance, slope steepness and slope orientation) with the point to estimate. In both tests, resulting LOO errors are higher than those provided by MLRK for all months. The best performance of MLRK could be firstly explained by the fact that the elevation is not the leading forcing factor for precipitation distribution in Norway and the integration of further geographical parameters into model equation improves the representation of the spatial variability of precipitation over the domain. Secondly, the uneven station distribution over Norway limits the spatial scales that could be actually resolved by the models. This station coverage is probably not suitable for interpolation approaches, such as LWLR, focusing on the reconstruction of precipitation gradients on a too local scale and it could lead to artifacts especially over remote areas characterizes by rough terrains and low data availability. MLRK turns out to be more robust over regions of sparse data coverage, since it performs the regression locally but considering a larger spatial scale, while the resolvable small-scale effect is then modeled by OK on regression residuals. \\

\subsection{Comparison with 2003-2016 HCLIM-AROME dataset}
\begin{figure}
\centering
\includegraphics[width=11cm,height=18cm]{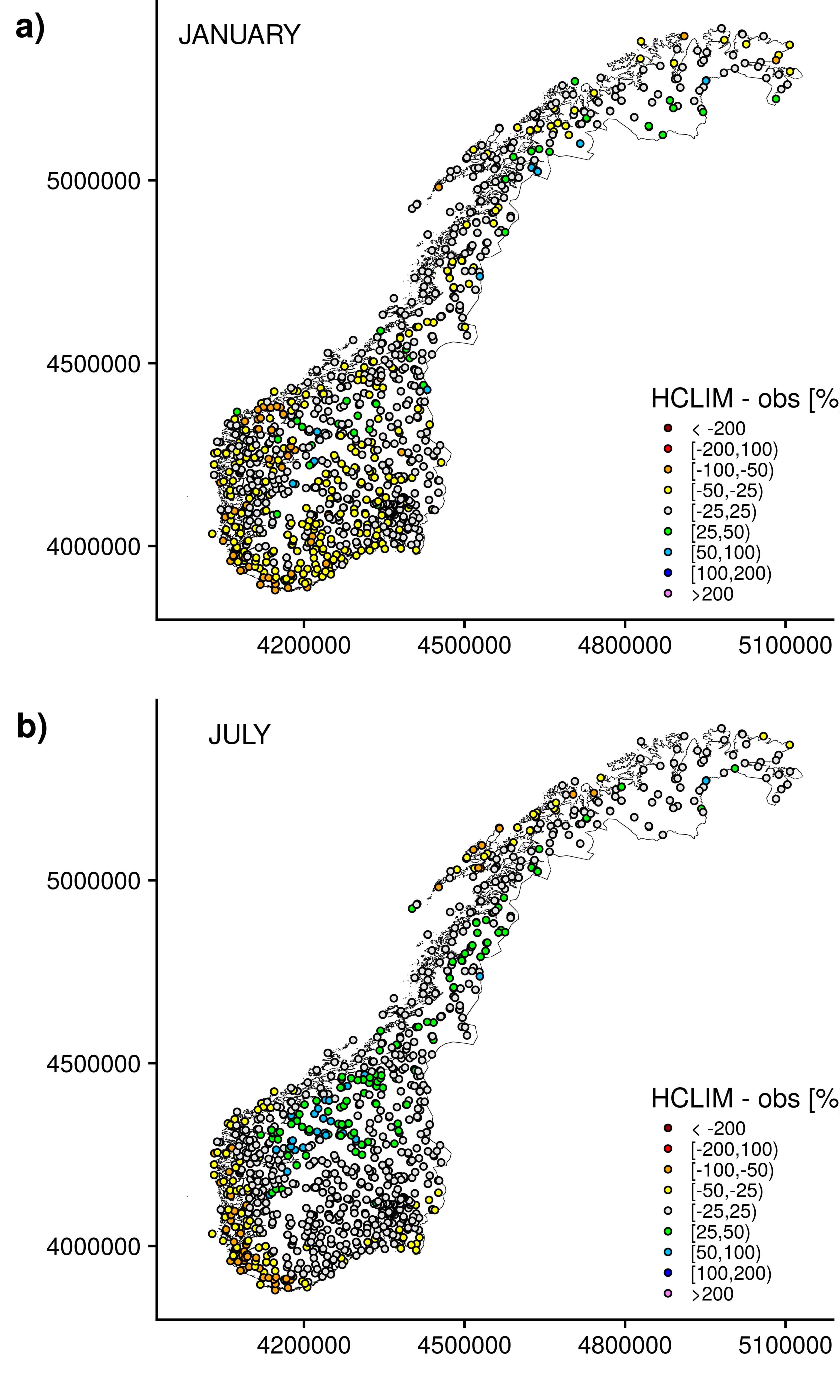}
\caption{Relative differences between HCLIM-AROME field and station normals for a) January and b) July.}
\label{fig:HCLIMbias}
\end{figure}

Robustness and limits of MLRK in properly capturing the relationships between rainfall and geographical features over Norway and in projecting them onto high-resolution grid were further assessed by performing a comparison with precipitation fields provided by HCLIM-AROME dataset. As already mentioned in section \ref{sub:HCLIM}, in order to perform comparison over the same time interval spanned by HCLIM-AROME series, 2003-2016 station normals were computed. Moreover, observation database was also redefined by including only the 2004 sites located inside area covered by HCLIM-AROME dataset.
In order to evaluate agreement between model results and station values, 2003-2016 monthly normals of all 1043 Norwegian stations were compared to values of the closest cells in the corresponding HCLIM-AROME grid. In table \ref{tab:HCLIM_MLRK} errors are reported together with those obtained by applying MLRK on the same station database of 2003-2016 monthly normals. 
\begin{table}[h]
\centering
\begin{tabular}{|l|lll|lll|}
\hline
&\multicolumn{3}{c|}{HCLIM-AROME} & \multicolumn{3}{c|}{MLRK} \\ \hline
MONTH &BIAS& MAE & RMSE & BIAS  & MAE & RMSE   \\ \hline
\multicolumn{1}{|c|}{1}&-21.5&26.5&38.1&0.4& 15.0& 23.1\\
\multicolumn{1}{|c|}{2}&-10.3&17.2&24.5&0.2&11.9&18.0\\
\multicolumn{1}{|c|}{3}&-12.6&20.0&28.3&0.3&12.1&18.8\\
\multicolumn{1}{|c|}{4}&-4.0&15.4&20.9&0.2&9.3&14.1\\
\multicolumn{1}{|c|}{5}&-2.5&14.6&19.5&0.0&8.4&12.2\\
\multicolumn{1}{|c|}{6}&-2.3&17.2&22.3&0.0&7.8&11.0\\
\multicolumn{1}{|c|}{7}&-1.0&20.4&27.2&0.1&8.3&11.7\\
\multicolumn{1}{|c|}{8}&-8.4&22.8&30.3&0.0&9.9&14.3\\
\multicolumn{1}{|c|}{9}&-21.2&28.5&41.0&0.0&13.7&22.3\\
\multicolumn{1}{|c|}{10}&-24.2&28.8&38.9&0.2&13.6&21.6\\
\multicolumn{1}{|c|}{11}&-23.6&31.5&44.3&0.4&15.7&24.7\\
\multicolumn{1}{|c|}{12}&-23.2&30.3&43.7&0.4&17.7&28.2\\
\hline
\end{tabular}
\caption{Accuracy of HCLIM-AROME monthly climatologies evaluated at station locations and of MLRK reconstruction of 2003-2016 station normals in LOO approach. In both cases validation subset is composed by 1043 Norwegian stations, values are expressed in mm and BIAS is defined as difference between model and observation.}
\label{tab:HCLIM_MLRK}
\end{table}
HCLIM-AROME field shows greatest errors in all months and a systematically negative bias, especially in winter season, corresponding to a general underestimation of station normals. This tendency is better represented in figure \ref{fig:HCLIMbias} showing the relative differences between HCLIM-AROME field and observed normals at Norwegian station sites for January and July. HCLIM-AROME monthly normals are lower than station values especially in coastal proximity, while in inland rainfall values at station points are generally overestimated by numerical model.
 
Despite possible bias in precipitation estimates for certain areas, pluviometric gradients depicted in HCLIM-AROME grids over regions without stations are likely to be more robust than those provided by statistical model since they are not affected by variation in data availability. Moreover, when observations are used to improve numerical outcomes by means of kriging-based distribution of station residuals described in section (HCLIM+KR), the resulting LOO reconstruction errors decrease significantly in all months and the bias is almost removed (table \ref{tab:HCLIMKR}). This error improvement suggests that the combination of resolved rainfall gradients of HCLIM-AROME fields with observation corrections could compensate both the uneven coverage of the station network and numerical model biases, thus providing more accurate climatologies over Norway.

\begin{table}[h]
\centering
 \begin{tabular}{||c c c c c||} 
 \hline
 MONTH & BIAS & MAE & RMSE  & \(R^2\) \\ [0.5ex] 
 \hline\hline
 1 & 0.3 & 12.9 & 19.4  & 0.9\\ 
 2 & 0.2 & 10.4 & 15.7 & 0.9\\
 3 & 0.2 & 10.5 & 16.3 & 0.9\\
 4 & 0.1 & 8.1 & 12.3 & 0.9 \\
 5 & 0.1 & 7.7 & 10.9 & 0.9\\ 
 6 & 0.0 & 7.6 & 10.4 & 0.8\\ 
 7 & 0.1 & 8.7 & 12.2 & 0.8\\ 
 8 & 0.1 & 9.5 & 13.2 & 0.9\\
 9 & 0.2 & 11.6 & 18.5 & 0.9 \\
 10 & 0.2 & 11.8 & 18.3 & 0.9\\ 
 11 & 0.3 & 13.7 & 21.8 & 0.9\\  
 12 & 0.3 & 16.0 & 25.0 & 0.9 \\ 
 \hline
\end{tabular}
\caption{Accuracy of HCLIM+KR monthly climatologies obtained from the leave-one-out validation for the 1043 stations in Norway. All values are expressed in mm and BIAS is defined as difference between simulation and observation.}
\label{tab:HCLIMKR}
\end{table}

\begin{figure}[h]
\centering
\includegraphics[width=11cm,height=18cm]{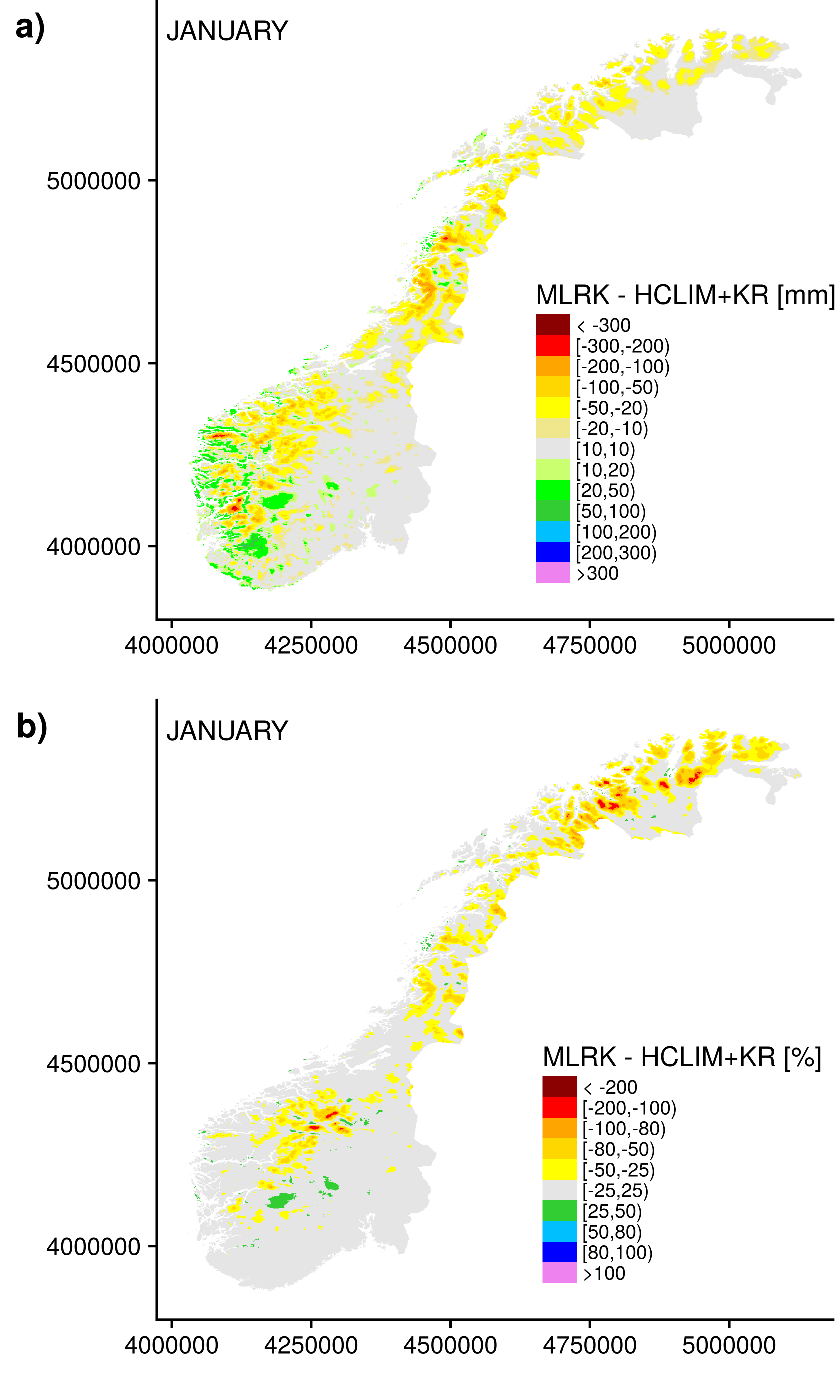}
\caption{Distribution of a) absolute and b) relative differences between MLRK and HCLIM+KR fields for January.}
\label{fig:modelBIAS1}
\end{figure}

\begin{figure}
\centering
\includegraphics[width=11cm,height=18cm]{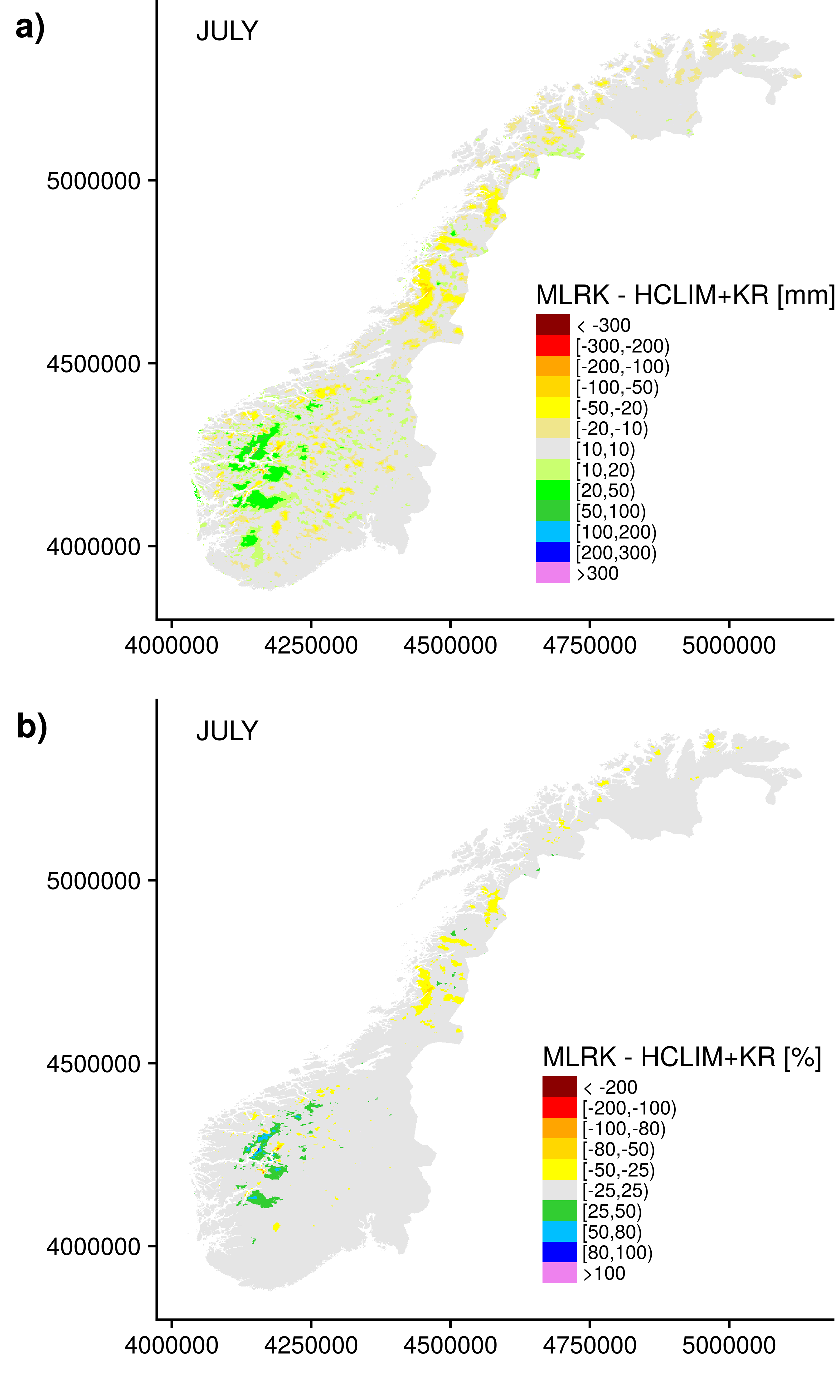}
\caption{Distribution of a) absolute and b) relative differences between MLRK and HCLIM+KR fields for July.}
\label{fig:modelBIAS7}
\end{figure}

\begin{figure}
\centering
\includegraphics[width=11cm,height=11cm]{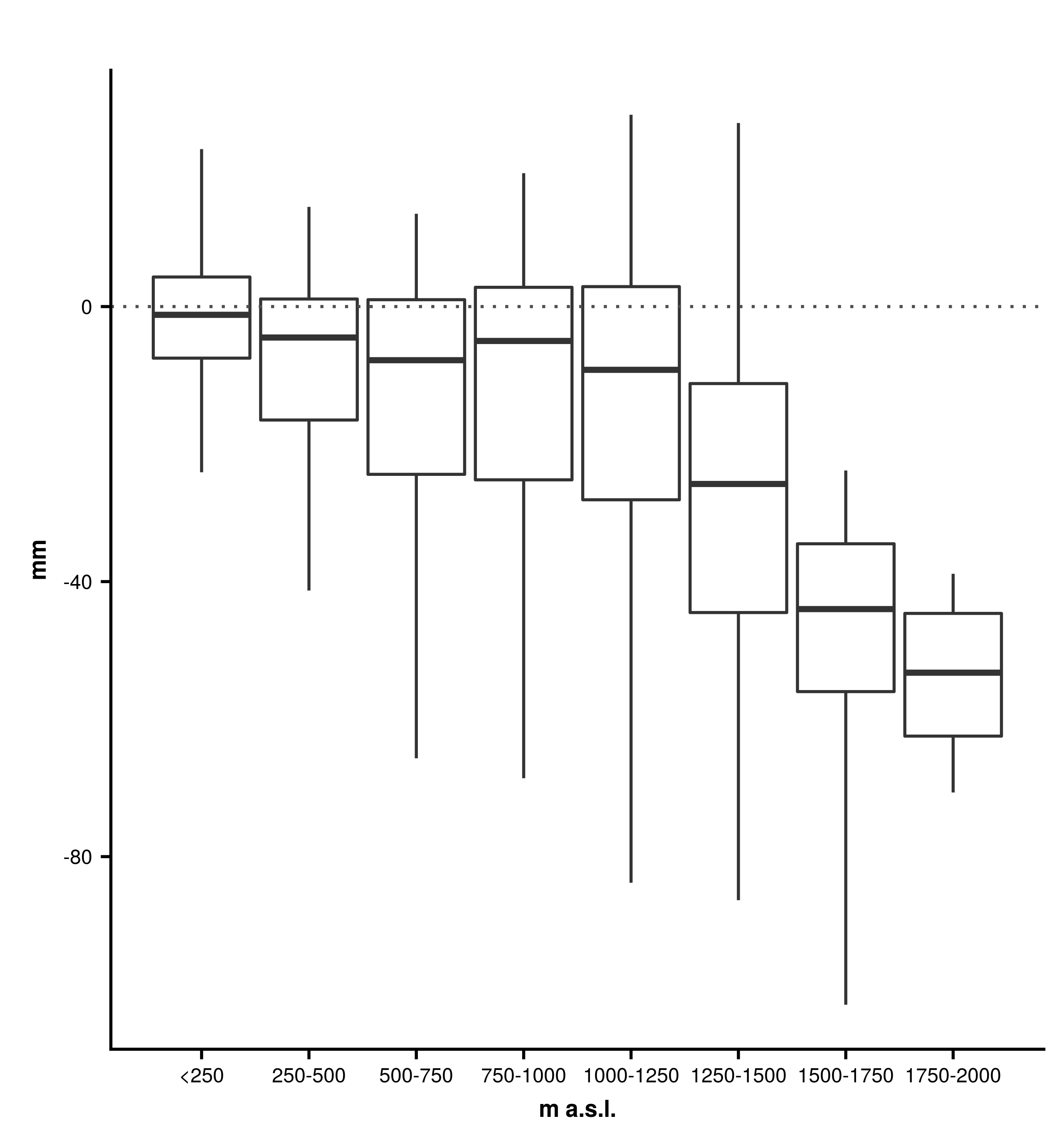}
\caption{Absolute bias between MLRK and HCLIM+KR precipitation normals for January clustered by elevation ranges.}
\label{fig:modelBIASelev}
\end{figure}

\begin{figure}
\centering
\includegraphics[width=11cm,height=11cm]{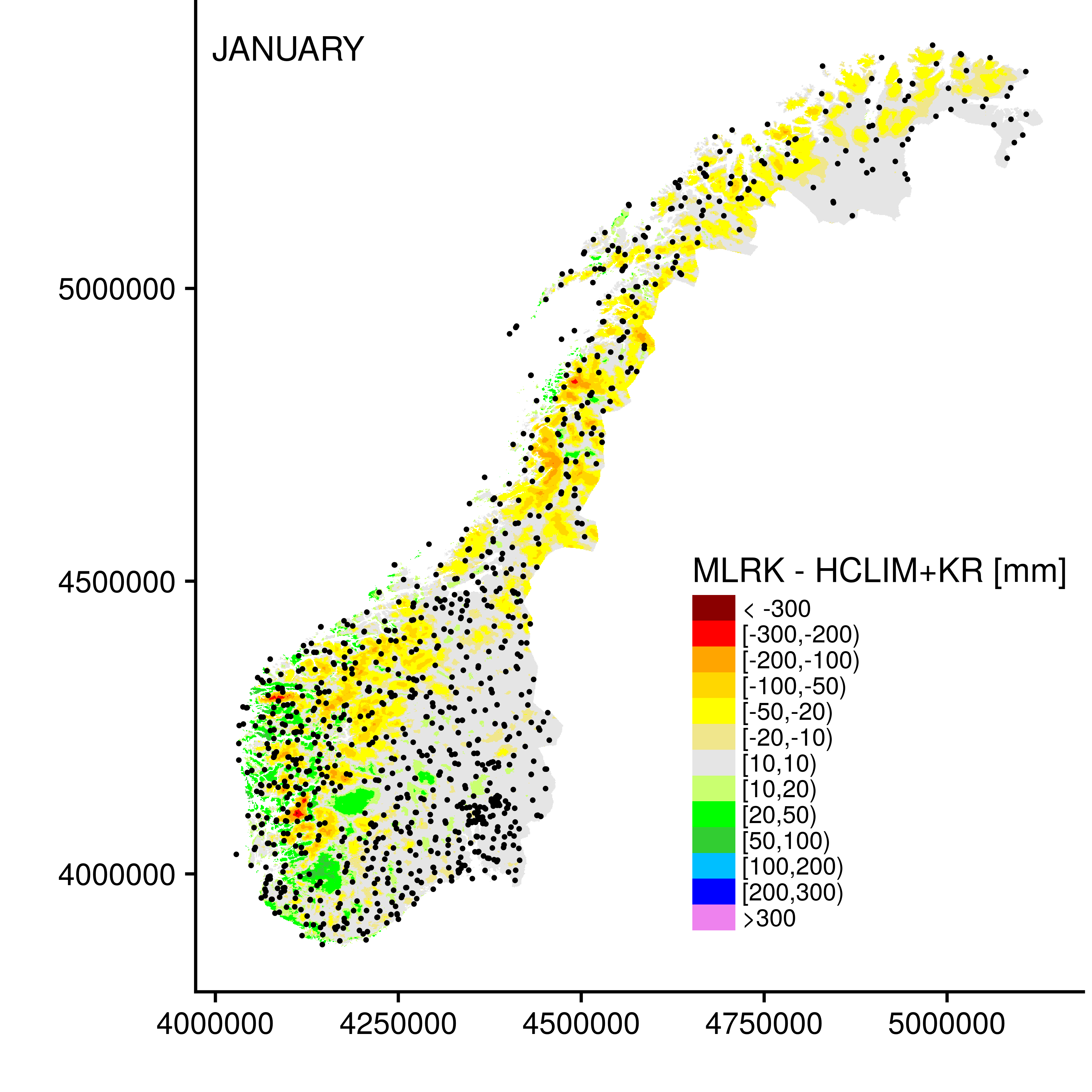}
\caption{Distribution stations overlapped to absolute differences between MLRK and HCLIM+KR fields for January.}
\label{fig:modelBIASstaz}
\end{figure}

In addition to station point comparison, MLRK and HCLIM+KR differences were evaluated also at grid cell level. In figures \ref{fig:modelBIAS1} and \ref{fig:modelBIAS7} the discrepancies between precipitation distributions provided by the two approaches for January and July are shown in both absolute and relative terms. While in July differences mainly concerned very limited portions of inner areas located in the highlands of Central and Southern Norway and range between -50\% and +50\%, in January disagreement is more evident and distributed on larger portions all over the  domain. Besides systematically lower HCLIM+KR values along West-Southern coast but with most deviations within 25\%, the greatest differences concern Southern inner highlands and mountain regions along Northern coasts, where MLRK reconstructs lower precipitation values and deviations from HCLIM+KR field overcome 80\% for points at the highest elevations. In order to further assess model differences, we clustered grid cells for classes of elevation and we calculated the boxplot of model discrepancies for January (figure \ref{fig:modelBIASelev}). The boxplot points out the MLRK tendency to provide lower precipitation normals than HCLIM+KR, especially at high-elevated grid cells. This behavior could be due to the difficulties of MLRK in extrapolating the actual precipitation-geographical relationship over complex regions where very few stations are available and mainly located at low elevation and/or, especially in Northern Norway, in coastal proximity. The role of data distribution is even more clear if station points are superimposed on the map of the model absolute differences for January (figure \ref{fig:modelBIASstaz}) proving that MLRK and HCLIM+KR discrepancies increase with decreasing station density, especially over reliefs. In addition, the variogram range (i.e. the maximum distance within which stations are spatially correlated) resulting from the station residuals of the multi-linear regression is generally below 20 km, while it is greater than 80 km if station residuals from HCLIM-AROME field are considered. The influence of observation residuals in correcting background field decreases more rapidly with distance in MLRK approach and emphasizes its difficulties to deal with remote regions.\\
The comparison, even if performed on a 14-year period of reference only, allowed to assess that the statistical scheme based on MLRK provides reasonable precipitation patterns over Norway. However, due to the uneven data coverage of the domain MLRK accuracy could significantly decrease over areas with very low station availability where actual pluviometric gradients could not be fully captured by regression. The combination of precipitation fields from a numerical model with precipitation normals derived from station observations could represent a promising approach to enhance the reconstruction of precipitation distribution also over Norwegian regions with very poor data coverage.

\section{1981-2010 monthly precipitation climatologies over Norway}

\begin{figure}[h]
\centering
\includegraphics[width=16cm,height=16cm]{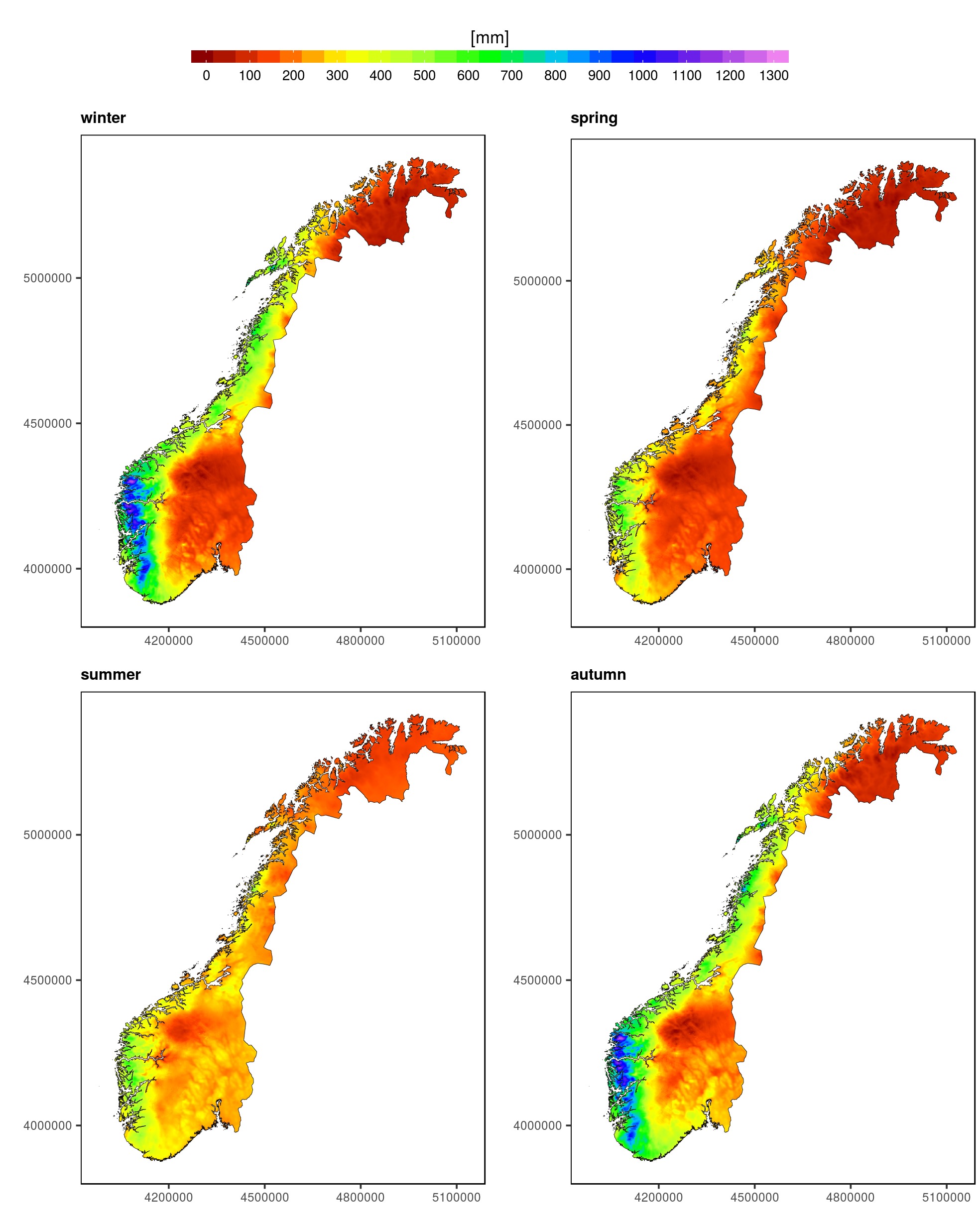}
\caption{Seasonal distribution of 1981-2010 precipitation normals over Norway gridded onto 1 km resolution by MLRK.}
\label{fig:seasonCLIM}
\end{figure}

\begin{figure}[h]
\centering
\includegraphics[width=16cm,height=16cm]{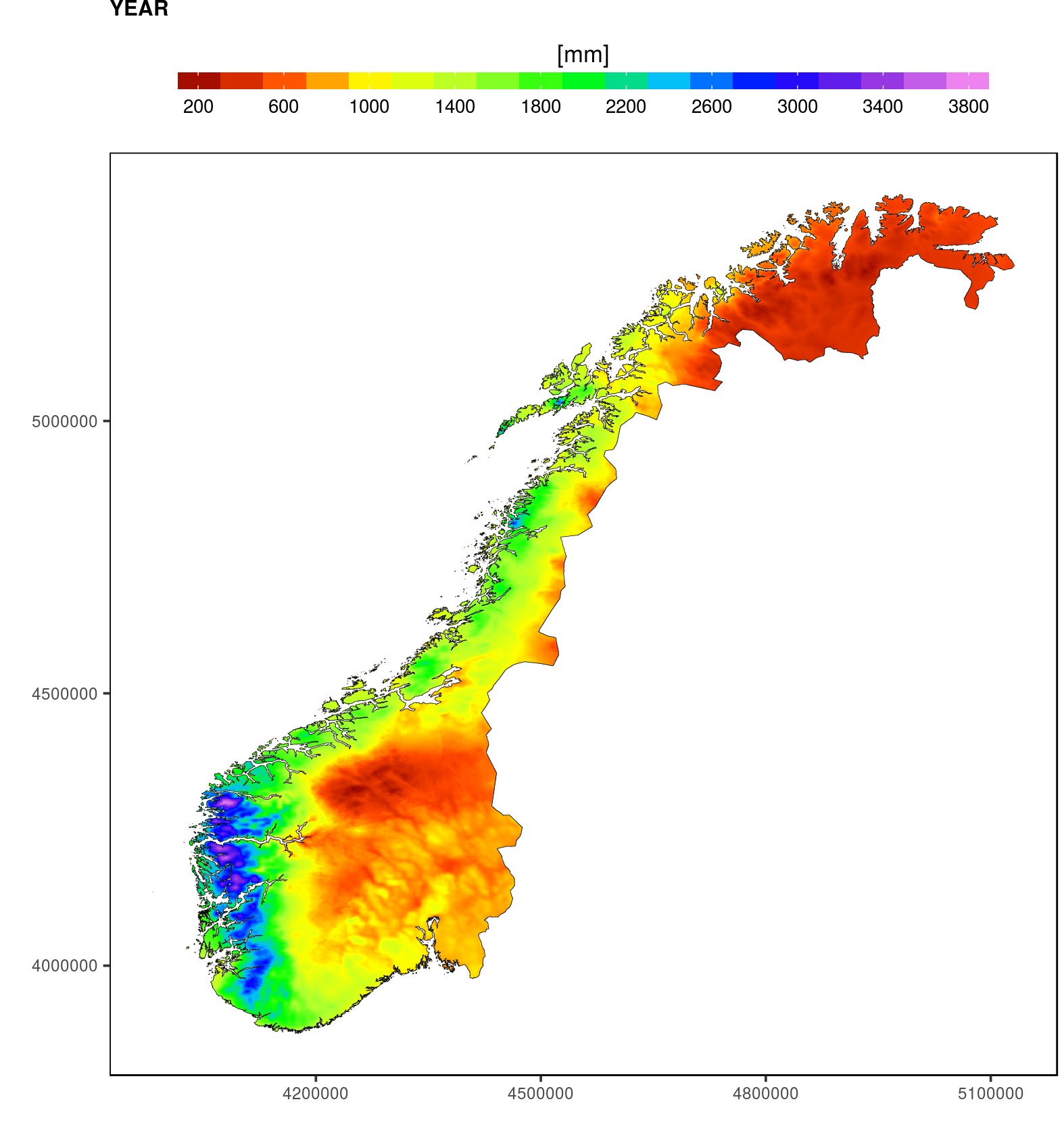}
\caption{Annual distribution of 1981-2010 precipitation normals over Norway gridded onto 1 km resolution by MLRK.}
\label{fig:annualCLIM}
\end{figure}

\begin{figure}[h]

\includegraphics[width=20cm,height=16cm]{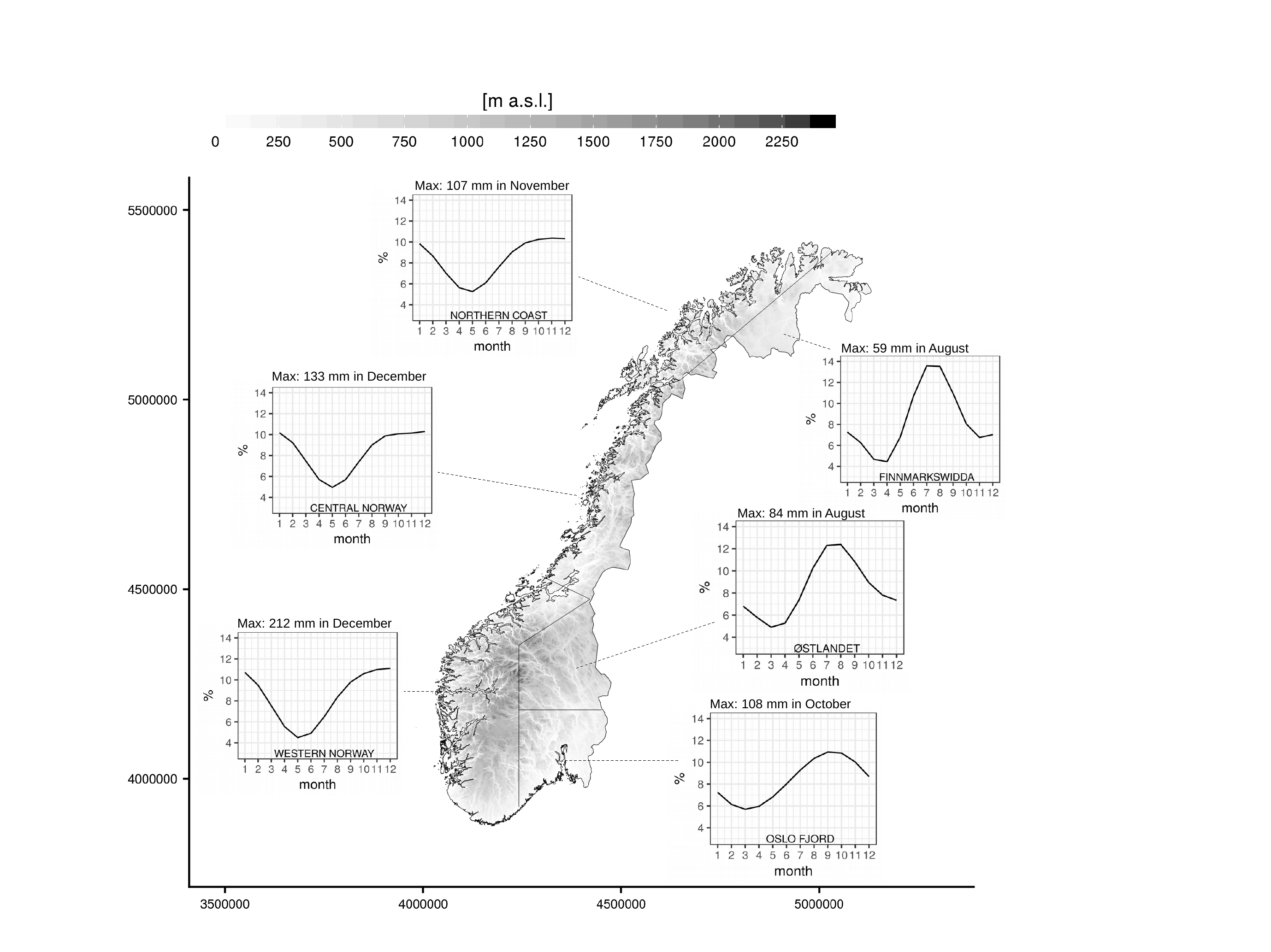}
\caption{Average yearly precipitation cycles of main Norwegian subregions. Values on y-axis are expressed as relative monthly contributions to total annual precipitation.}
\label{fig:seasonality}
\end{figure}

In figures \ref{fig:seasonCLIM} and \ref{fig:annualCLIM} MLRK seasonal and annual 1981-2010 Norwegian precipitation climatologies at 1 km resolution are shown. A west-to-east gradient is evident throughout the whole year with the highest precipitation values occurring along the westernmost part of Norway and a clear transition to continental regime towards inland regions. The coastal mountain ridge all along western Norway acts as an important orographic barrier for moist air masses from the sea which are quickly lifted up and loss their water vapor over a very short distance from the coast. For this reason, the coastal regions experience the greatest precipitation over the whole year, especially in south-western Norway where the highest reliefs occurs and annual maxima of more than 3500 mm are reconstructed. Areas featuring the lowest annual precipitation are located in the northernmost mountainous part of the country and east of the mountain ridge in the South with total amounts generally below 400 mm. On the seasonal scale, wintertime is the wettest period with an average precipitation value of 302 mm and maxima of more than 1200 mm over the south-western mountain ridge, especially in the area close to Sognefjorden. Spring turns to be, with about 190 mm as areal mean, the driest season when a large portion of inland regions, especially the upper Gudbrandsdalen, features minimum precipitation amounts and seasonal totals below 100 mm. It is noteworthy that, as explained in previous sections, these areas, especially in the North, are also the least covered by observations and gridded values are mostly extrapolated in MLRK by using distant stations located in different climatic areas leading in some cases to unreliable regression equations which emphasize dry conditions and provide, even if for a very limited number of points, monthly values slightly negative or very close to zero. Since these outcomes are intrinsic in the modeling structure and could be removed only by increasing data coverage, grid points with negative occurrences are retained and used to identify the areas mostly affected by model uncertainty, while their negative estimations are set to zero in the cumulate rainfall computation. Further information about Norwegian climate could be retrieved by considering average yearly precipitation cycles of main Norwegian subregions. As shown in figure \ref{fig:seasonality}, the most remarkable distinction, also in terms of annual precipitation pattern, regards coastal and inland regions. In particular, Norwegian areas which are close to sea experience the highest relative monthly contributions to annual precipitation in winter and autumn and the lowest ones in late spring and this annual pattern turns out to be almost constant with latitude. On the contrary, in inner regions, for which  Finnmarkswidda and Østlandet are considered, the main contributions to annual precipitation are provided by summer months while minima are shifted to spring and early spring. Moreover, yearly rainfall cycle of inlands shows a greater variation with latitude with an increase in summer relative contributions of about 2\% of the total moving Northern Norway. Oslo Fjord represents an intermediate area with a smoother annual cycle showing the lowest precipitation in early spring and the maximum contributions in late summer when convective phenomena are dominant.

\section{Conclusions}
A new version of 1981-2010 monthly precipitation climatologies at 1 km resolution has been produced over Norway.
The climatologies are based on ECA\&D and KDVH daily observation databases over the Scandinavian peninsula.
As a preliminary step, the station data have been quality-checked and a daily-gap filling procedure has been applied.
The interpolation of 1981-2010 station monthly normals has been performed by means of a multi-linear local regression kriging (MLRK). The statistical interpolation method combines a local regression of station precipitation normals \textit{versus} several geographical predictors with a kriging interpolation of the station regression residuals.
The geographical parameters considered are elevation, latitude, longitude and distance from the sea.
The MLRK performances have been evaluated by studying the summary statistics derived from a leave-one-out cross-validation approach of the station monthly normals. All the Norwegian stations have been considered as the basis for the cross-validation.
The monthly errors, in terms of MAE, range between 7.1 mm in May and 16.0 mm in January, while \(R^2\) is above 0.8 in all months.
MLRK have been compared to other statistical interpolation approaches, such as a local weighted linear precipitation-elevation regression (LWLR) and a regression kriging considering only elevation as precipitation predictor.
MLRK turns out to be preferable than the other interpolation approaches over Norway. In fact, the spatial scale considered by MLRK fits better with the uneven station distribution of the domain and the inclusion of more than one regression predictor allows to improve the model estimations.\\
MLRK gridded precipitation have been also compared to HCLIM-AROME fields derived from a 2003-2016 reanalysis driven regional climate model with a resolution of 2.5 km.
The numerical model provides realistic precipitation fields, that are independent from the in-situ observations. On the other hand, HCLIM-AROME estimates show higher biases than MLRK, especially over coastal locations.
A new combined interpolation approach (HCLIM+KR) has been implemented and evaluated over the 2003-2016 period. 
In this procedure the HCLIM-AROME fields are used as the background for the kriging interpolation of the station residuals so that the gridded precipitation is estimated by means of both observations and numerical output.
If compared to the MLRK cross-validation results, this integrated product leads to a significant decrease of validation errors and it represents a valid approach to increase the accuracy of the gridded precipitation climatologies over Norway, provided the availability of HCLIM-AROME fields covering longer periods than 2003-2016.\\
According to the 1981-2010 monthly precipitation climatologies computed by MLRK over Norway, the annual and seasonal maps show a strong west-to-south gradient occurring between the wet mountainous coastal regions and the continental climate of the outback.
The annual totals range between about 4000 mm over the west coast and less than 300 mm in the southern inner areas and in Finnmarkswidda. 
The comparison of the mean annual cycles of six Norwegian subdomains proves the opposite climate regime occurring between coasts and inlands but it allows also to depict more in detail the specific subregional features of annual precipitation patterns over the country.

\clearpage
\pagebreak

\bibliography{biblio1}

\begin{thebibliography}{11}
\providecommand{\natexlab}[1]{#1}
\expandafter\ifx\csname urlstyle\endcsname\relax
  \providecommand{\doi}[1]{doi:\discretionary{}{}{}#1}\else
  \providecommand{\doi}{doi:\discretionary{}{}{}\begingroup
  \urlstyle{rm}\Url}\fi

\bibitem[{\textit{Crespi et~al.}(2018)\textit{Crespi, Brunetti, Lentini, and
  Maugeri}}]{cres18}
Crespi, A., M.~Brunetti, G.~Lentini, and M.~Maugeri (2018), 1961–1990
  high-resolution monthly precipitation climatologies for italy,
  \textit{International Journal of Climatology}, \textit{38}(2), 878--895.

\bibitem[{\textit{Daly et~al.}(1994)\textit{Daly, Neilson, and
  Phillips}}]{DALY94}
Daly, C., R.~P. Neilson, and D.~L. Phillips (1994), A statistical-topographic
  model for mapping climatological precipitation over mountainous terrain,
  \textit{Journal of Applied Meteorology}, \textit{33}(2), 140--158.

\bibitem[{\textit{Daly et~al.}(2002)\textit{Daly, Gibson, Taylor, Johnson, and
  Pasteris}}]{DALY02}
Daly, C., W.~P. Gibson, G.~H. Taylor, G.~L. Johnson, and P.~Pasteris (2002), A
  knowledge-based approach to the statistical mapping of climate,
  \textit{Climate Research}, \textit{22}(2), 99--113.

\bibitem[{\textit{Foresti and Pozdnoukhov}(2011)}]{FORESTI2011}
Foresti, L., and A.~Pozdnoukhov (2011), Exploration of alpine orographic
  precipitation patterns with radar image processing and clustering techniques,
  \textit{Meteorological Applications}, \textit{19}(4), 407--419.

\bibitem[{\textit{Goovaerts}(2000)}]{GOOV2000}
Goovaerts, P. (2000), Geostatistical approaches for incorporating elevation
  into the spatial interpolation of rainfall, \textit{Journal of Hydrology},
  \textit{228}(1), 113 -- 129.

\bibitem[{\textit{Hofstra and New}(2009)}]{HOF09}
Hofstra, N., and M.~New (2009), Spatial variability in correlation decay
  distance and influence on angular-distance weighting interpolation of daily
  precipitation over europe, \textit{International Journal of Climatology},
  \textit{29}(12), 1872--1880.

\bibitem[{\textit{Hofstra et~al.}(2008)\textit{Hofstra, Haylock, New, Jones,
  and Frei}}]{hofstra2008comparison}
Hofstra, N., M.~Haylock, M.~New, P.~Jones, and C.~Frei (2008), Comparison of
  six methods for the interpolation of daily, european climate data,
  \textit{Journal of Geophysical Research: Atmospheres}, \textit{113}(D21).

\bibitem[{\textit{Lind et~al.}(2016)\textit{Lind, Lindstedt, Kjellström, and
  Jones}}]{Lind2016HCLIM}
Lind, P., D.~Lindstedt, E.~Kjellström, and C.~Jones (2016), Spatial and
  temporal characteristics of summer precipitation over central europe in a
  suite of high-resolution climate models, \textit{Journal of Climate},
  \textit{29}(10), 3501--3518, \doi{10.1175/JCLI-D-15-0463.1}.

\bibitem[{\textit{Masson and Frei}(2014)}]{MASSFREI14}
Masson, D., and C.~Frei (2014), Spatial analysis of precipitation in a
  high-mountain region: exploring methods with multi-scale topographic
  predictors and circulation types, \textit{Hydrology and Earth System
  Sciences}, \textit{18}(11), 4543--4563.

\bibitem[{\textit{Masson et~al.}(2013)\textit{Masson, Le~Moigne, Martin,
  Faroux, Alias, Alkama, Belamari, Barbu, Boone, Bouyssel, Brousseau, Brun,
  Calvet, Carrer, Decharme, Delire, Donier, Essaouini, Gibelin, Giordani,
  Habets, Jidane, Kerdraon, Kourzeneva, Lafaysse, Lafont, Lebeaupin~Brossier,
  Lemonsu, Mahfouf, Marguinaud, Mokhtari, Morin, Pigeon, Salgado, Seity,
  Taillefer, Tanguy, Tulet, Vincendon, Vionnet, and
  Voldoire}}]{Masson2013SURFEX}
Masson, V., P.~Le~Moigne, E.~Martin, S.~Faroux, A.~Alias, R.~Alkama,
  S.~Belamari, A.~Barbu, A.~Boone, F.~Bouyssel, P.~Brousseau, E.~Brun, J.-C.
  Calvet, D.~Carrer, B.~Decharme, C.~Delire, S.~Donier, K.~Essaouini, A.-L.
  Gibelin, H.~Giordani, F.~Habets, M.~Jidane, G.~Kerdraon, E.~Kourzeneva,
  M.~Lafaysse, S.~Lafont, C.~Lebeaupin~Brossier, A.~Lemonsu, J.-F. Mahfouf,
  P.~Marguinaud, M.~Mokhtari, S.~Morin, G.~Pigeon, R.~Salgado, Y.~Seity,
  F.~Taillefer, G.~Tanguy, P.~Tulet, B.~Vincendon, V.~Vionnet, and A.~Voldoire
  (2013), The surfexv7.2 land and ocean surface platform for coupled or offline
  simulation of earth surface variables and fluxes, \textit{Geoscientific Model
  Development}, \textit{6}(4), 929--960, \doi{10.5194/gmd-6-929-2013}.

\bibitem[{\textit{Seity et~al.}(2011)\textit{Seity, Brousseau, Malardel, Hello,
  Bénard, Bouttier, Lac, and Masson}}]{Seity2001Arome}
Seity, Y., P.~Brousseau, S.~Malardel, G.~Hello, P.~Bénard, F.~Bouttier,
  C.~Lac, and V.~Masson (2011), The arome-france convective-scale operational
  model, \textit{Monthly Weather Review}, \textit{139}(3), 976--991,
  \doi{10.1175/2010MWR3425.1}.

\end{thebibliography}

\clearpage
\pagebreak
\end{document}